# Migration of Accreting Planets and Black Holes in Disks

JT Laune[1], Rixin Li (李日新)[1,2,4], and Dong Lai[1,3]
[1] Department of Astronomy, Cornell Center for Astrophysics and Planetary Science, Cornell University, Ithaca, NY 14853, USA
[2] Department of Astronomy, University of California, Berkeley, CA 94720, USA
[3] Tsung-Dao Lee Institute, Shanghai Jiao Tong University, 200240 Shanghai, People's Republic of China

## Abstract

Nascent planets are thought to lose angular momentum (AM) to the gaseous protoplanetary disk via gravitational interactions, leading to inward migration. A similar migration process also applies to stellar-mass black holes (BHs) embedded in the disks of active galactic nuclei. However, AM exchange via accretion onto the planet/BH may strongly influence the migration torque. In this study, we perform 2D global hydrodynamic simulations of an accreting planet/BH embedded in a disk, where AM exchange between the planet/BH and disk via gravity, accretion, pressure, and viscosity are considered. When accretion is turned off, we recover the linear estimate for Type I migration torque. However, for the two planet masses we investigated with our accreting simulations, we find outward migration due to the positive AM deposited onto the accreting body by the disk gas. Our simulations achieve the global steady state for the transport of mass and AM: The mass and AM fluxes are constant across the disk except for jumps ($\Delta \dot{M}$ and $\Delta \dot{J}$) at the planet's location, and the jumps match the accretion rate and torque on the planet. Our findings suggest that caution is needed when applying the standard results of disk migration to accreting planets and BHs.

*Unified Astronomy Thesaurus concepts:* Protoplanetary disks (1300); Accretion (14); Planetary migration (2206); Active galactic nuclei (16); Black holes (162)

## 1. Introduction

Nascent planets exchange orbital angular momentum (AM) and energy with their natal protoplanetary disks (PPDs), leading to the evolution of their orbits. This process is collectively referred to as planet migration and has been extensively studied over the last four decades (P. Goldreich & S. Tremaine 1979, 1980; D. N. C. Lin & J. Papaloizou 1979; W. R. Ward 1986, 1997; for a review, see S.-J. Paardekooper et al. 2023). A similar migration process also applies to stars or stellar-mass black holes (BHs) embedded in the disks of active galactic nuclei (AGN, e.g., P. Artymowicz et al. 1993; Y. Levin 2007; B. McKernan et al. 2012; J. M. Bellovary et al. 2016).

Migration is mainly divided into two regimes, Type I (linear) and Type II (nonlinear, gap-forming). In the standard picture of viscous Type I migration, a planet perturbs the gas disk and the (slightly asymmetric) gravitational back-reaction on the planet facilitates the AM exchange. In almost all hydrodynamical studies of migration, the planet is modeled as a softened point-mass potential, and the gravitational torque between the disk and planet is measured. The speed and direction of Type I migration have received much attention (H. Tanaka et al. 2002; S.-J. Paardekooper et al. 2011; W. Kley & R. P. Nelson 2012), and depend on various factors, such as the shape of the corotation region (F. S. Masset 2001; F. S. Masset & G. I. Ogilvie 2004; S. J. Paardekooper & J. C. B. Papaloizou 2009a, 2009b) and thermal structure of the disk (S.-J. Paardekooper & G. Mellema 2006; C. Baruteau & F. Masset 2008; S.-J. Paardekooper & J. C. B. Papaloizou 2008; W. Kley & A. Crida 2008; F. S. Masset & J. Casoli 2009, 2010; M. A. Jimenez & F. S. Masset 2017). Many of these studies find inward migration of the planet.

More recent migration studies have incorporated realistic gas and dust physics to determine how they affect migration. For example, the inclusion of thermal diffusion and radiative heating of the disk by the planet has led to the discovery and investigation of thermal torques (E. Lega et al. 2014; P. Benítez-Llambay et al. 2015; O. Chrenko et al. 2017; H. Eklund & F. S. Masset 2017; F. S. Masset 2017; A. M. Hankla et al. 2020; R. O. Chametla & F. S. Masset 2021). For low-mass, luminous planets, thermal torques may reverse migration and drive eccentricity and inclination, whereas for cold planets, they accelerate inward migration. In these studies, the AM exchange still principally occurs through gravity.

However, planets accrete disk gas that carries AM, leading to extra AM exchange besides gravitational torque and a change in the total AM budget of the planet. This effect is particularly relevant for giant planet formation, during runaway core accretion, when a rocky core of around $10\,M_\oplus$ is able to accrete all of the gas supplied to it (J. B. Pollack et al. 1996; M. Ikoma et al. 2000; R. R. Rafikov 2006; Y. Hori & M. Ikoma 2011; B. A. Ayliffe & M. R. Bate 2012; A.-M. A. Piso & A. N. Youdin 2014).

Treating planetary migration and accretion simultaneously is difficult due to the disparate length scales. Except a few, most studies treat the processes separately. W. Kley et al. (2001) and G. D'Angelo et al. (2002, 2003) use a nested grid to achieve high resolution around the planet and empty the cells within a certain radius around the planet to simulate accretion. Accretion only affects the torque on the planet by modifying the gas density close to the planet and thereby the gravitational force it feels. C. Dürmann & W. Kley (2017) follow a planet on a live orbit and model accretion by reducing the surface density

---
[4] 51 Pegasi b Fellow.

in the vicinity of the planet. These studies typically find that accretion slows down migration.

Accretion-driven AM exchange is also of great interest for stellar-mass BHs embedded in an AGN disk around a supermassive BH. Stellar-mass BHs in the disk can migrate like planets and might get trapped in some regions where the migration torque vanishes (B. McKernan et al. 2012; J. M. Bellovary et al. 2016). These migration traps are a promising potential site for binary BH formation (A. Secunda et al. 2019; H. Tagawa et al. 2020; J. Li et al. 2022). The role accretion plays in migration could therefore have important implications for binary BH formation.

The case of an accreting body (planet or BH) in a disk is like an extreme mass ratio binary. Circumbinary accretion disks have long been thought to sap AM from the binary and lead to its hardening, but recent simulations of circumbinary accretion (D. J. Munoz et al. 2019; M. S. L. Moody et al. 2019; D. J. Muñoz et al. 2020; D. J. D'Orazio & P. C. Duffell 2021; J. Zrake et al. 2021; H.-Y. Wang et al. 2023; P. C. Duffell et al. 2024; for a review, see D. Lai & D. J. Muñoz 2023), which track the AM transferred to the binary through accretion and gravity, find that under some conditions AM can flow into the binary, leading to orbital expansion. The analogous behavior for a planet/BH in a disk is outward migration. However, even after accounting for accretion, some studies of circumbinary accretion with different parameters have continued to find inward migration (e.g., P. C. Duffell et al. 2020; R. M. Heath & C. J. Nixon 2020; A. J. Dittmann & G. Ryan 2022, 2024). The expected behavior for planetary systems is unclear.

In this paper, we study the migration of an accreting "planet" (which can be a planet or a BH) in a viscous, isothermal disk. We treat the planet as an absorbing sphere and track the AM accreted by the planet, the pressure and viscous torques, and the AM exchange by gravity. Using an iterative procedure, we evolve the disk self-consistently so that it reaches a global steady state, i.e., the mass or AM flux is constant across the disk except for a jump at the planet's location, with the magnitude of the jump equal to the mass-accretion rate or torque on the planet. In Section 2, we describe the numerical methods we employ for the planet and the disk. In Sections 3 and 4, we present our results for nonaccreting and accreting planets, respectively. We conclude in Section 5.

## 2. Methods

To investigate the effects of accretion on planet migration, we use the code Athena++ (J. M. Stone et al. 2020) in a rotating annulus. Section 2.1 describes our simulation setup for modeling the global disk flow and the flow around the planet. Sections 2.2 and 2.3 describe our methods for calculating disk mass and AM fluxes and planetary accretion and torques, respectively. In Section 2.4, we describe how we treat the boundaries and how we iterate the simulations to find self-consistent global solutions.

### 2.1. Numerical Setup

We consider a planet of mass $m_p$ on a circular orbit around a star of mass $M$ with semimajor axis (SMA) $a_p$ and angular velocity $\mathbf{\Omega}_p = \Omega_p \hat{z} = \sqrt{GM/a_p^3}\,\hat{z}$. It is embedded in a thin gaseous disk. We work in a frame centered on the star, rotating at $\mathbf{\Omega}_p$, so that the planet remains fixed. Let $\mathbf{r} = (r, \varphi)$ denote the position and $\mathbf{v} = (v_r, v_\varphi)$ the gas velocity (in the rotating frame).

We model the gaseous disk in an annulus, $\mathcal{D}$, centered on the star between $r_{in} = 0.4a_p$ and $r_{out} = 1.6a_p$. We solve the viscous compressible Navier–Stokes equations in the rotating frame in 2D:

$$\frac{\partial \Sigma}{\partial t} + \nabla \cdot (\Sigma \mathbf{v}) = 0, \quad (1)$$

$$\frac{\partial (\Sigma \mathbf{v})}{\partial t} + \nabla \cdot (\Sigma \mathbf{v}\mathbf{v} + P\mathbf{I} + \Sigma \nu \tilde{\mathbf{T}})$$
$$= -2\Sigma \mathbf{\Omega}_p \times \mathbf{v} - \Sigma \mathbf{\Omega}_p \times (\mathbf{\Omega}_p \times \mathbf{r})$$
$$- \Sigma (\nabla \Phi + \nabla \Phi_p + \nabla \Phi_{ind}), \quad (2)$$

where $\Sigma$ is the surface (vertically integrated) density of the gas; $\nu$ the kinematic viscosity and $\tilde{T}$ the specific viscous stress tensor; $P$ the pressure; and $\Phi$, $\Phi_p$, and $\Phi_{ind}$ are the potentials for the star's gravity, the planet's gravity, and the indirect term from using the star-centered frame. The pressure is assumed to be barotropic and globally isothermal, $P = c_s^2 \Sigma$, where the sound speed, $c_s$, is constant. The disk scale height at the planet's location is $H = c_s/\Omega_p$ (with aspect ratio $h = H/a_p$). We assume the kinematic viscosity is constant. The equivalent Shakura–Sunyaev parameter at $r = a_p$ is $\alpha = \nu \Omega_p/c_s^2$.

We solve Equations (1) and (2) in $\mathcal{D}$ in a cylindrical $r$–$\varphi$ geometry, using the HLLE Riemann flux solver, the second-order van Leer time integrator, and the piecewise linear method for spatial reconstruction. The domain is discretized into a base grid with resolution $N_r \times N_\varphi$, and we utilize Athena++'s static mesh refinement (SMR) to refine the grid around the planet by a factor of $2^{N_{SMR}}$ in each direction.

### 2.2. Disk Accretion Profiles and Steady State

The (inward) disk mass flux through the ring at radius $r$, $\dot{M}(r)$, is given by the integral

$$\dot{M}(r) = \oint (-rv_r)\Sigma d\varphi. \quad (3)$$

The AM advected through the ring is

$$\dot{J}_{adv}(r) = \oint (-r^2 \Sigma v_r)(v_\varphi + r\Omega_p)d\varphi. \quad (4)$$

The AM diffused through the ring by viscosity is

$$\dot{J}_{vis}(r) = \oint (-r^3 \nu \Sigma)\left[\frac{\partial}{\partial r}\left(\frac{v_\varphi}{r}\right) + \frac{1}{r^2}\frac{\partial v_r}{\partial \varphi}\right]d\varphi. \quad (5)$$

The torque density exerted by the planet's gravity at the ring is

$$\frac{dT_g}{dr} = \oint (-r\Sigma)\frac{\partial \Phi_p}{\partial \varphi}d\varphi, \quad (6)$$

so that the gravitational torque is given by

$$T_g^{>r}(r) = \int_r^{r_{out}} \frac{dT_g}{dr}dr. \quad (7)$$

The total (inward) AM flux at radius $r$ is thus

$$\dot{J}(r) = \dot{J}_{adv} - \dot{J}_{vis} - T_g^{>r}. \quad (8)$$

See R. Miranda et al. (2017) for a derivation of these quantities.

An unperturbed, axisymmetric disk in steady state has radially constant mass flux and AM flux, so we write $\dot{M}(r) = \dot{M}_{out}$ (the mass supply rate at the outer disk boundary) and $\dot{J}(r) = \dot{J}_{\text{out}}$, and define the specific AM "eigenvalue" as

$$j_{\text{out}} = \frac{\dot{J}_{\text{out}}}{\dot{M}_{\text{out}}}. \tag{9}$$

The steady-state surface density profile is

$$\Sigma_{\text{out}}(r) = \frac{\dot{M}_{\text{out}}}{3\pi\nu}\left(1 - \frac{j_{\text{out}}}{j}\right), \tag{10}$$

where $j(r) = v_\varphi r + \Omega_p r^2 \simeq \sqrt{GMr}$ is the specific AM of the gas. The steady-state velocity profile (in the rotating frame) is $\mathbf{v}_{\text{out}} = (v_{r,\text{out}}, v_{\varphi,\text{out}})$, where the radial velocity profile is

$$v_{r,\text{out}}(r) = -\frac{\dot{M}_{\text{out}}}{2\pi r \Sigma_{\text{out}}} = -\frac{3\nu}{2r}\left(1 - \frac{j_{\text{out}}}{j}\right)^{-1}, \tag{11}$$

and the azimuthal velocity profile is

$$v_{\varphi,\text{out}}(r) = v_K - r\Omega_p + \frac{c_s^2 j_{\text{out}}}{4 j v_K}\left(1 - \frac{j_{\text{out}}}{j}\right)^{-1}, \tag{12}$$

where $v_K(r) = (GM/r)^{1/2}$ is the Keplerian velocity. The $v_\varphi$ profile is non-Keplerian due to the pressure support.

If a nonaccreting planet perturbs the disk, the time-averaged $\dot{M}$ profile remains constant across the disk, and the above profiles provide a self-consistent description of the disk away from the planet's location.[5] On the contrary, the perturbation from an accreting planet leads to discontinuities in the mass flux and AM flux at $r = a_p$, with the time-averaged $\dot{M}$ and $\dot{J}$ piecewise constant inside and outside $a_p$. We denote the disk values inside the planet as $\dot{M}(r < a_p) = \dot{M}_{\text{in}}$ (i.e., the accretion rate through the inner boundary) and $\dot{J}(r < a_p) = \dot{J}_{\text{in}}$, and we set $j_{\text{in}} = \dot{J}_{\text{in}}/\dot{M}_{\text{in}}$. The corresponding surface density and velocity are $\Sigma_{\text{in}}$ and $\mathbf{v}_{\text{in}} = (v_{r,\text{in}}, v_{\varphi,\text{in}})$, calculated with equations analogous to Equations (10)–(12). Note that while $\dot{M}_{\text{out}}$ is an arbitrary constant that sets the disk density scale, $\dot{M}_{\text{in}}$, $j_{\text{out}}$, and $j_{\text{in}}$ must be determined self-consistently. Without loss of generality, we set $j_{\text{in}} = 0$ in our simulations; this is equivalent to assuming that our computational inner boundary $r_{\text{in}}$ is much larger than the true (physical) inner boundary of the disk. We also define the step in $\dot{M}$ to be

$$\Delta \dot{M} = \dot{M}_{\text{out}} - \dot{M}_{\text{in}}, \tag{13}$$

and the step in $\dot{J}$ to be

$$\Delta \dot{J} = \dot{J}_{\text{out}} - \dot{J}_{\text{in}}. \tag{14}$$

In a (time-averaged) steady state, we expect $\Delta \dot{M}$ to be equal to the planet's mass-accretion rate and $\Delta \dot{J}$ the net torque on the planet.

### 2.3. Planet Orbital Evolution, Accretion, and Torque Calculation

The key parameters for an accreting planet are its mass-accretion rate, $\dot{m}_p$, and its total AM exchange (i.e., torque) with the disk, $\dot{J}_p$, arising from gravitational forces, accreted momentum, pressure forces, and viscous forces. If the radius of the planet is small, only a negligible part of $\dot{J}_p$ will go to the planet's spin; the rest will go to its orbital AM. The key migration parameter is then $l_p = \dot{J}_p/(\dot{m}_p a_p^2 \Omega_p)$. The planet's

SMA evolves as

$$\frac{\dot{a}_p}{a_p} = 2(l_p - 1)\frac{\dot{m}_p}{m_p} - \frac{\dot{M}_{\text{in}}}{M}. \tag{15}$$

Since $m_p \ll M$, the second term is usually negligible compared to the first. If $l_p > 1$, the orbit will expand rather than contract.

We model the planet's gravity as a softened point-mass potential,

$$\Phi_p = -\frac{Gm_p}{\sqrt{|\mathbf{r} - \mathbf{r}_p|^2 + b^2}}, \tag{16}$$

where $b$ is the softening length. The gravitational torque exerted on the planet is

$$\dot{J}_{p,g} = \int_{\mathcal{D}} (-a_p \Sigma)(\hat{\varphi}_p \cdot \nabla_p \Phi_p) r dr d\varphi. \tag{17}$$

The total torque on a nonaccreting planet is $\dot{J}_p = \dot{J}_{p,g}$.

An accreting planet is treated as an absorbing sphere with sink radius, $r_s$, and evaluation radius, $r_e$ (see R. Li & D. Lai 2022, 2023, 2024). At every time step, in the region $|\mathbf{r} - \mathbf{r}_p| < r_s$, we set $\Sigma/(\dot{M}_{\text{out}}/3\pi\nu) = 0.01$, $\mathbf{v} = 0$, and $P/(c_s^2 \dot{M}_{\text{out}}/3\pi\nu) = 0.01$. We evaluate the planet's gas accretion-related quantities through a circle of radius $r_e$ centered on the planet, i.e., the circle $\mathcal{C}_e$ defined by the locus $|\mathbf{r} - \mathbf{r}_p| = r_e$. The evaluation radius $r_e$ is chosen to be $r_e = r_s + \delta_r \sqrt{2}$, where $\delta_r = 1.2/(2^{N_{\text{SMR}}} N_r)$ is the grid resolution at the highest level of SMR ($N_{\text{SMR}} = 5$ and $N_r = 64$; see also Section 2.5). This choice ensures that the interpolation of hydrodynamic quantities at $\mathcal{C}_e$ does not utilize any sink cells.

The infinitesimal mass-accretion rate through an arc of $\mathcal{C}_e$ with length $dl$ and unit normal $\hat{\mathbf{n}}$ (measured from $\mathbf{r}_p$) is

$$d\dot{m}_p = -\Sigma \mathbf{v} \cdot \hat{\mathbf{n}} dl. \tag{18}$$

The mass-accretion rate onto the planet, $\dot{m}_p$, is the path integral (along $\mathcal{C}_e$):

$$\dot{m}_p = \oint_{\mathcal{C}_e} d\dot{m}_p. \tag{19}$$

The torque on the planet from the accreted gas momentum is

$$\dot{J}_{p,a} = \oint_{\mathcal{C}_e} a_p \hat{\varphi}_p \cdot (\mathbf{v} + \Omega_p \hat{\mathbf{z}} \times \mathbf{r}) d\dot{m}_p. \tag{20}$$

The torque due to pressure is

$$\dot{J}_{p,p} = \oint_{\mathcal{C}_e} (-a_p P) \hat{\varphi}_p \cdot \hat{\mathbf{n}} dl. \tag{21}$$

The viscous torque is

$$\dot{J}_{p,v} = \oint_{\mathcal{C}_e} a_p \Sigma \nu \hat{\mathbf{n}} \cdot \tilde{\mathbf{T}} dl. \tag{22}$$

To numerically calculate the path integrals (Equations (19–22)), we evaluate the integrands at 100 evenly spaced points on $\mathcal{C}_e$, $\mathbf{r}_i$, and sum the results using a nearest-neighbor interpolation scheme for $\Sigma(\mathbf{r}_i)$ and $\mathbf{v}(\mathbf{r}_i)$.[6] Equations (19)–(21) are evaluated 100 times every orbit and saved. $\dot{J}_{p,v}$ is calculated in post-processing every orbit and is very small, and hence we neglect it from future analysis. The total torque on the accreting

---

[5] This assumes that the net torque on the planet is much less than $\dot{M} j(a_p) = \dot{M}_{\text{out}} \sqrt{GMa_p}$. This condition is satisfied for nonaccreting planets.

[6] Testing with more points does not make a difference.

Table 1
The Six Different Simulations in This Paper and Their Results

| Run | $\frac{m_p}{M}$ (1) | $\frac{r_s}{r_H}$ (2) | $\frac{r_s}{a_p}$ (3) | $\frac{\dot{M}_{in}}{\dot{M}_{out}}$ (4) | $\frac{j_{out}}{a_p^2\Omega_p}$ (5) | $\frac{\dot{m}_p}{\dot{M}_{out}}$ (6) | $\frac{\dot{J}_p}{\dot{M}_{out}a_p^2\Omega_p}$ (7) | $l_p$ (8) | $\frac{\dot{a}_p}{(\dot{M}_{out}/m_p)a_p}$ (9) |
|---|---|---|---|---|---|---|---|---|---|
| In  | $10^{-3}$ | ⋯ | ⋯ | ⋯ | ⋯ | ⋯ | ⋯ | ⋯ | −0.0189 |
| Ia  | $10^{-3}$ | 0.05 | 0.00347 | 0.299 | 0.742 | 0.635 | 0.688 | 1.08 | 0.106 |
| I   | $10^{-3}$ | 0.1 | 0.00693 | 0.287 | 0.737 | 0.708 | 0.748 | 1.06 | 0.0788 |
| Ib  | $10^{-3}$ | 0.2 | 0.0139 | 0.259 | 0.766 | 0.825 | 0.839 | 1.02 | 0.0276 |
| IIn | $10^{-4}$ | ⋯ | ⋯ | ⋯ | ⋯ | ⋯ | ⋯ | ⋯ | $-7.56 \times 10^{-4}$ |
| II  | $10^{-4}$ | 0.1 | 0.00693 | 0.561 | 0.455 | 0.488 | 0.494 | 1.01 | 0.0121 |

**Notes.** Columns: (1) planet-to-star mass ratio; (2) sink radius (in units of the planet's Hill radius, $r_H$); (3) sink radius (in units of the planet's SMA, $a_p$); (4) the inner mass-accretion rate reached through iteration (Equation (29); see Section 2.4); (5) the outer AM flux eigenvalue reached through iteration (Equation (30)), in units of $a_p^2\Omega_p = \sqrt{GMa_p}$, the specific AM of the planet; (6) planet mass-accretion rate (Equation (19)); (7) planet total AM exchange rate (Equation (23)); (8) the planet accretion "eigenvalue," $l_p = \dot{J}_p/(\dot{m}_p a_p^2 \Omega_p)$; (9) orbital evolution of the planet (Equation (15)). Runs In and IIn are simulations with a nonaccreting planet, for which we only list the $\dot{a}_p$ value (see text); the other runs are simulations with an accreting planet. Run Ia utilizes six levels of static mesh refinement as opposed to five for the others (see Section 2.1).

planet is

$$\dot{J}_p = \dot{J}_{p,g} + \dot{J}_{p,a} + \dot{J}_{p,p}. \quad (23)$$

Mass conservation in the simulation domain should ensure that $\dot{m}_p = \Delta \dot{M}$. Similarly, AM conservation should ensure $\dot{J}_p = \Delta \dot{J}$. The quantities $\Delta \dot{M}$ and $\Delta \dot{J}$ are calculated as the flux through $\mathcal{D}$ (the integrals in Equations (3)–(5) and (7)), whereas $\dot{m}_p$ and $\dot{J}_p$ are calculated as flux through $\mathcal{C}_e$. Because of this difference, it is an independent test of our numerical scheme that $\dot{m}_p = \Delta \dot{M}$ and $\dot{J}_p = \Delta \dot{J}$.

### 2.4. Boundary Conditions, Initial Conditions, and Iterations

In the inner and outer ghost zones ($r < 0.4a_p$ and $r > 1.6a_p$), we set the hydrodynamic variables to their steady-state values, $\Sigma_{in}, \mathbf{v}_{in}$ and $\Sigma_{out}, \mathbf{v}_{out}$, respectively (see Section 2.2). The azimuthal boundary condition is periodic.

In the annular region between $r = 1.4a_p$ to $1.6a_p$, which we denote $\mathcal{W}_{out}$ and refer to as the "outer wave damping zone," we damp the gas quantities to their steady-state values with the formulae

$$\left(\frac{\partial \Sigma}{\partial t}\right)_{d,out} = -\zeta(x_{out})\frac{\Sigma - \Sigma_{out}}{T_d}, \quad (24)$$

$$\left(\frac{\partial \Sigma \mathbf{v}}{\partial t}\right)_{d,out} = -\zeta(x_{out})\frac{\Sigma \mathbf{v} - \Sigma_{out}\mathbf{v}_{out}}{T_d}. \quad (25)$$

In the "inner wave damping zone," $\mathcal{W}_{in}$, which extends from $r = 0.4a_p$ to $0.6a_p$, we damp the gas quantities with formulae analogous to Equations (24)–(25) and parameter $x_{in}$. The wave damping zones $\mathcal{W}_{in}$ and $\mathcal{W}_{out}$ model an infinite steady-state disk outside of $\mathcal{D}$ and let waves travel out of $\mathcal{D}$ to "infinity" rather than reflecting at the boundaries. Here, $\zeta$ is a tapering function:

$$\zeta(x) = \begin{cases} 1 - 6x^2 + 6x^3 & 0 \leq x \leq \frac{1}{2} \\ 2(1-x)^3 & \frac{1}{2} < x \leq 1, \end{cases} \quad (26)$$

and $x_{in}$ and $x_{out}$ are defined to be

$$x_{in} = \frac{r - 0.4}{0.2}, \quad (27)$$

$$x_{out} = 1 - \frac{r - 1.4}{0.2}. \quad (28)$$

We initialize the disk with a nonaccreting planet with $j_{out} = j_{in} = 0$ and $\dot{M}_{out} = \dot{M}_{in}$, i.e., $\Sigma = \Sigma_{out} = \dot{M}_{out}/3\pi\nu$ is radially constant, $v_r = v_{r,out} = -3\nu/2r$, and $v_\varphi = v_{\varphi,out}$ is Keplerian. We integrate the disk for 100 orbits, damping the gas variables to their initial values in $\mathcal{W}_{in}$ and $\mathcal{W}_{out}$, and then measure $\dot{J}_p$, $\dot{m}_p$, $\dot{M}(r)$, and $\dot{J}(r)$.

Then, we restart the simulation with an accreting planet and integrate an additional 30 orbits. The wave damping zones are initially inconsistent with an accreting planet due to the discontinuities in $\dot{M}$ and $\dot{J}$ at the planet's location. To find self-consistent values for $\dot{M}_{in}, j_{in}$ and $\dot{M}_{out}, j_{out}$ in $\mathcal{W}_{in}$ and $\mathcal{W}_{out}$ (and the ghost cells), we use an iterative procedure. At the end of the first integration, we measure the values of $\dot{M}_{in}, \dot{M}_{out}$ and $\dot{J}_{in}, \dot{J}_{out}$ in the regions $r \in [0.6, 1]a_p$ and $[1, 1.4]a_p$, respectively. We denote these measured quantities with the subscript "old" and denote quantities we use for the next simulation with "new." These data give us initial guesses for the accretion rate $\Delta\dot{M}_{old} = \dot{M}_{out,old} - \dot{M}_{in,old}$ and torque $\Delta\dot{J}_{old} = \dot{J}_{out,old} - \dot{J}_{in,old}$. We assume no AM is transferred through the inner boundary (or to the central star) and keep $j_{in,new} = 0$. We also keep $\dot{M}_{out,new} = \dot{M}_{out,old}$. Then, we set

$$\frac{\dot{M}_{in,new}}{\dot{M}_{out,new}} = 1 - \frac{\Delta\dot{M}_{old}}{\dot{M}_{out,old}} \quad (29)$$

and

$$j_{out,new} = \frac{\Delta\dot{J}_{old}}{\dot{M}_{out,old}}. \quad (30)$$

Together, Equations (29) and (30) define $\Sigma_{in,new}, \Sigma_{out,new}, \mathbf{v}_{in,new}$, and $\mathbf{v}_{out,new}$. We restart the integration with these new values in $\mathcal{W}_{in}$ and $\mathcal{W}_{out}$ and allow the solution to viscously relax. We repeat this procedure until $j_{out,new}$ and $\dot{M}_{in,new}$ converge and $\dot{M}(r)$ and $\dot{J}(r)$ exhibit step functions. We consider our iterative procedure converged whenever the changes in $\dot{M}$ and $\dot{J}$ at the wave damping zone boundaries are less than 5% of $\dot{M}_{out}$ and $\dot{J}_{out}$, respectively.

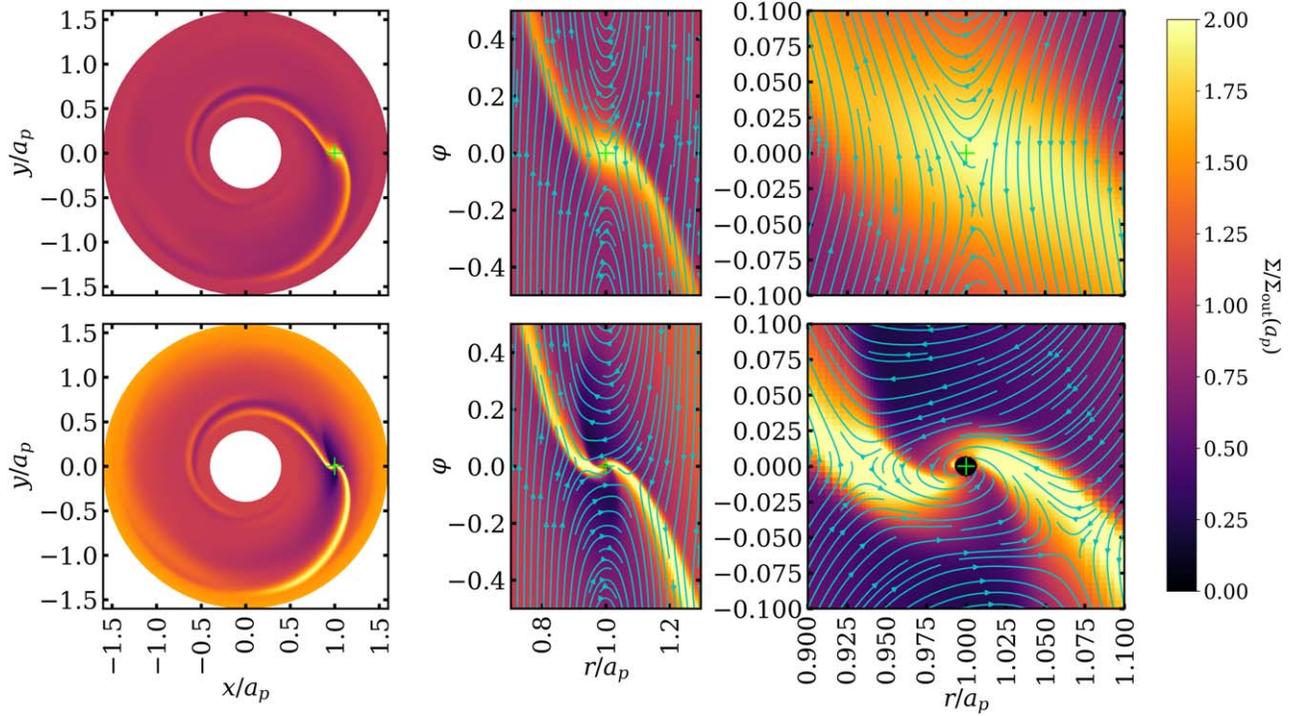

**Figure 1.** Left: the global disk surface density for the nonaccreting Run In (top) and the accreting Run I (bottom; see Table 1). The green "+" indicates the planet location. Middle: zoomed-in plots of the surface density in polar coordinates with stream plots indicating gas velocity. Right: same as the middle column, but zoomed closer to the planet.

### 2.5. Numerical Parameters

In this study, we perform six sets of simulations (see Table 1 for a summary). In all of them, we set $c_s = 0.07\Omega_p a_p$ (so that $h = 0.07$ at $r = a_p$) and $\nu = 5 \times 10^{-4} a_p^2 \Omega_p$ (corresponding to $\alpha = 0.1$ at $r = a_p$). We choose the damping timescale $T_d = 0.01\Omega_p^{-1}$ for the inner and outer wave damping zones.

We consider two planet masses, $q = m_p/M = 10^{-3}$ and $10^{-4}$, which we label Run I and II, respectively. We choose $q = 10^{-3}$ as the fiducial planetary mass ratio so that the Hill radius of the planet, $r_H = (q/3)^{1/3} a_p$, is equal to the scale height of the disk, justifying our (approximate) 2D treatment. The fiducial sink radius we use in Runs I and II is $r_s = 0.1 r_H$. To investigate the effect of our choice of $r_s$, we set up two variations of Run I. In Run Ia, we set $r_s = 0.05 r_H$, and in Run Ib, we set $r_s = 0.2 r_H$, all other parameters being equal. All accreting planets have their softening length $b = 0$.

For both planet masses, in addition to an accreting planet we also simulate a nonaccreting planet, referred to as Runs In and IIn. In Section 3, we discuss the choice of softening length for nonaccreting planets.

For the base grid, we choose $N_r = 64$ and $N_\varphi = 384$. All runs have $N_{SMR} = 5$, except for Run Ia, which has 6. The finest level covers a $0.02 a_p \times 0.02 a_p$ square centered on the planet. This enables us to resolve the sink radius by ~10 cells (introduced in Section 2.3) while evolving the entire disk with reasonable computational efficiency.

### 3. Results for Nonaccreting Planets

We simulate a nonaccreting planet (Runs In and IIn) for 100 orbits so that the flow reaches a quasi-steady state. Figure 1 (top panels) shows the final flow structure in the global disk and in the vicinity of the planet. We see that the planet launches spiral waves inside and outside its orbit. The gas undergoes horseshoe turns in the co-orbital region, approximately from $r/a_p = 0.9$ to 1.1. The width, $\Delta r_{cor}/a_p \approx 0.2$, is in good agreement with previous numerical and analytic studies, which find the width of the corotation region to be (S. J. Paardekooper & J. C. B. Papaloizou 2009a; C. W. Ormel 2013)

$$\frac{\Delta r_{cor}}{a_p} = \sqrt{\frac{q}{h}}, \qquad (31)$$

which gives $\Delta r_{cor}/a_p = 0.21$ for Run I.

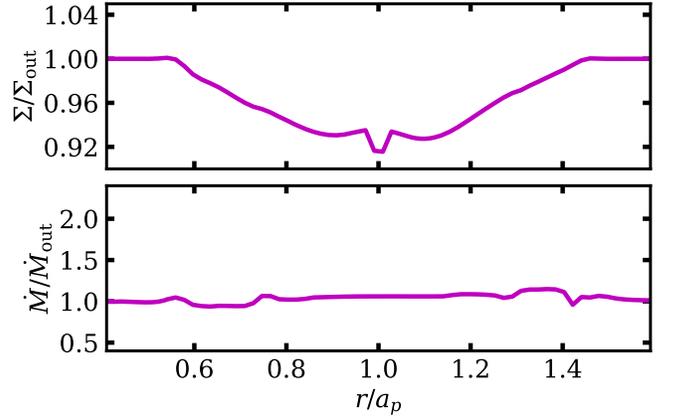

**Figure 2.** Disk profiles for the nonaccreting Run In. Top: the azimuthally averaged surface density profile, $\Sigma(r)$. Bottom: the mass flux through the disk, $\dot{M}(r)$ (Equation (3)).

In Figure 2, we show the disk profiles for the nonaccreting Run In. The gap in surface density, $\Sigma/\Sigma_{out}$, is shallow, ~90% (for Run IIn, with $q = 10^{-4}$, there is effectively no gap). The $\dot{M}$ profile is relatively flat, with $\dot{M}/\dot{M}_{out}$ approximately constant to within 15% (and about 3% in Run IIn). In Figure 3, we show the $\dot{j}$ component profiles for the disk.

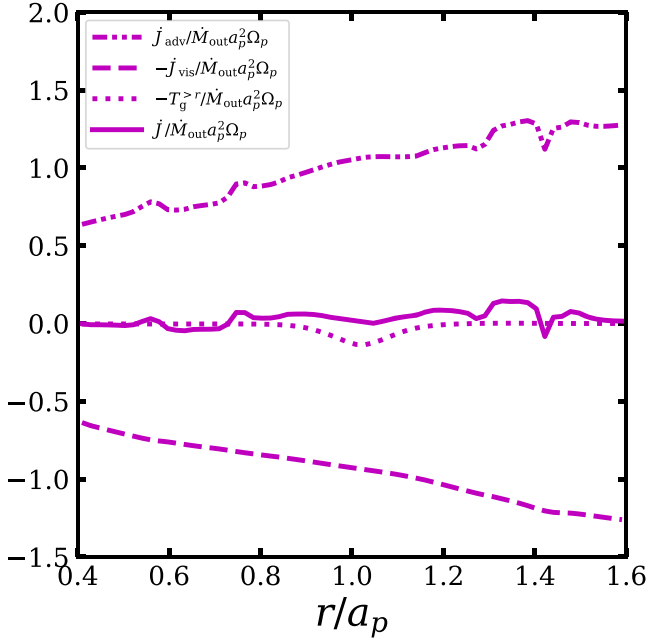

**Figure 3.** The components of AM flux through the disk for Run In. The dashed–dotted line is the advected AM flux, $\dot{J}_{adv}$ (Equation (4)). The dashed line is the viscous AM flux, $-\dot{J}_{vis}$ (Equation (5)). The dotted line is the torque density exerted by the planet's gravity, $-T_g^{>r}$ (Equation (7)). The solid line is the total AM flux, $\dot{J}$ (Equation (8)).

Because $\dot{J}_{adv}$ and $\dot{J}_{vis}$ cancel each other out, the total $\dot{J}$ profile is close to zero (consistent with our setup with $j_{in} = j_{out} = 0$), deviating from zero only by about $0.1(\dot{M}_{out}a_p^2\Omega_p)$. The large bumps in $\dot{M}$ and $\dot{J}$ near $r/a_p = 0.7$ and 1.3 correspond to the boundary of the first level of SMR in the simulation grid. The results are similar for Run IIn.

The standard linear gravitational torque for Type I migration is approximately given by (S.-J. Paardekooper et al. 2023)

$$\left(\frac{\dot{j}_{p,g}}{\Sigma_{out}a_p^4\Omega_p^2}\right)_{linear} \approx -\left(\frac{q}{h}\right)^2\left(\frac{\Sigma_p}{\Sigma_{out}}\right), \quad (32)$$

where $\Sigma_p$ is the surface density of the disk in the vicinity of the planet. For our disk model (constant $\nu$ and $j_{out} = 0$ for a nonaccreting planet), $\Sigma_p \simeq \Sigma_{out}$, and the linear estimates are $\dot{j}_{p,g}/(\Sigma_{out}a_p^4\Omega_p^2) = 2.0 \times 10^{-4}$ ($q = 10^{-3}$, Run In) and $2.0 \times 10^{-6}$ ($q = 10^{-4}$, Run IIn), respectively.

In previous studies, the softening length, $b$ (Equation (16)), is typically chosen to be a fraction of the scale height, $H$ (e.g., M. De Val-Borro et al. 2006; S.-J. Paardekooper et al. 2010). This suffices for the low-mass planet in Run IIn ($q = 10^{-4}$), for which we set $b/a_p = 0.6h = 0.042$. The torque we measure in the simulation is $\dot{j}_{p,g}/(\Sigma_{out}a_p^4\Omega_p^2) = -1.78 \times 10^{-6}$, in agreement with the linear result. The corresponding orbital evolution rate is $\dot{a}_p/a_p = -7.56 \times 10^{-4}\dot{M}_{out}/m_p$.

However, we find that the torque on the higher-mass planet in Run In is sensitive to our choice of $b$. In particular, for $b/a_p = 0.042$, the torque is positive, $\dot{j}_{p,g}/(\Sigma_{out}a_p^4\Omega_p^2) = 2.65 \times 10^{-4}$. The torque switches sign from positive to negative between $b/a_p = 0.07$ [$\dot{j}_{p,g}/(\Sigma_{out}a_p^4\Omega_p^2) = 1.80 \times 10^{-5}$] and $b/a_p = 0.08$ [$\dot{j}_{p,g}/(\Sigma_{out}a_p^4\Omega_p^2) = -1.11 \times 10^{-5}$]. We recover the linear result with $b/a_p = 0.1$, for which we find $\dot{j}_{p,g}/(\Sigma_{out}a_p^4\Omega_p^2) = -4.45 \times 10^{-5}$. The corresponding orbital evolution is $\dot{a}_p/a_p = -0.0189\dot{M}_{out}/m_p$.

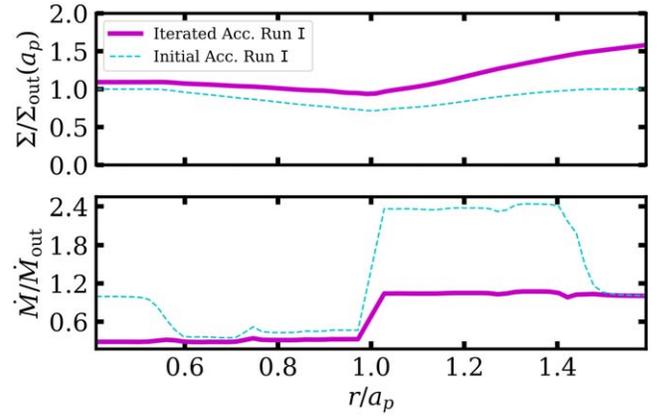

**Figure 4.** Same as Figure 2, but for an accreting planet in Run I. The magenta lines indicate the profiles from the iterated simulation (see Section 2.4); the dashed cyan lines indicate the profiles before iteration.

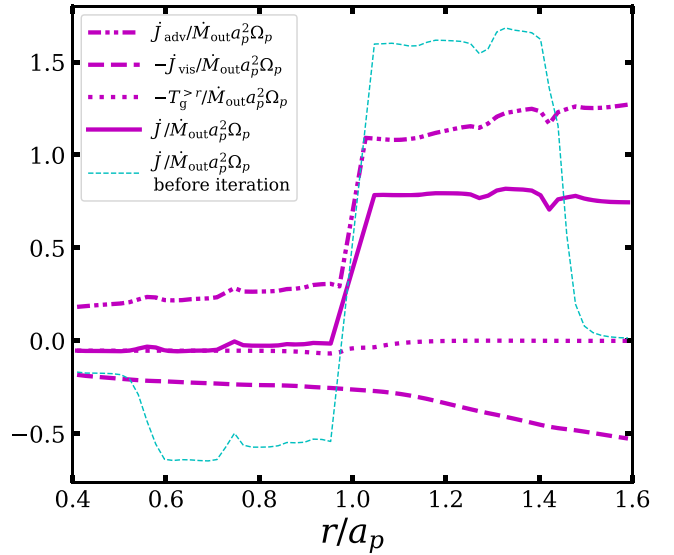

**Figure 5.** Same as Figure 3, but for an accreting planet in Run I. The magenta lines indicate the profiles from the iterated simulation. The dashed cyan line indicates the $\dot{J}$ profile before iteration.

## 4. Results for Accreting Planets

To simulate accreting planets, we restart the nonaccreting runs, turn on accretion, and integrate for 30 orbits. Then, we iterate using the procedure described in Section 2.4 until we reach self-consistent boundary conditions. The results are listed in the third and fourth columns of Table 1. Figure 1 compares the final steady-state flow structures between the case of an accreting planet (Run I; after iteration) and that of a nonaccreting planet (Run In; see also Section 3). We see that the region around the accreting planet is significantly more depleted outside of the spiral waves when compared to the area in the nonaccreting case. The corotation region in Run I extends approximately from $r/a_p = 0.85$ to 1.15, with a corotation width $\Delta r_{cor}/a_p \approx 0.3$ that is $\sim$50% larger than in Run In.

In Figure 4, we show the disk $\Sigma(r)$ and $\dot{M}(r)$ profiles for an accreting planet in Run I. The surface density $\Sigma(r)$ slopes upward outside of the planet, $r > a_p$, and is relatively flat inside, $r < a_p$. In Figure 5, we show the $\dot{J}$ component profiles after iteration and the $\dot{J}$ profile before iteration. Before iteration, $\dot{M}(r)$ and $\dot{J}(r)$ both "find" different steady-state (piecewise constant) values outside of the wave damping zones in the region $r = 0.6 a_p$–$1.4 a_p$. After iteration, $\dot{M}(r)$ and $\dot{J}(r)$ both attain a step-function behavior, and a global steady state is achieved.

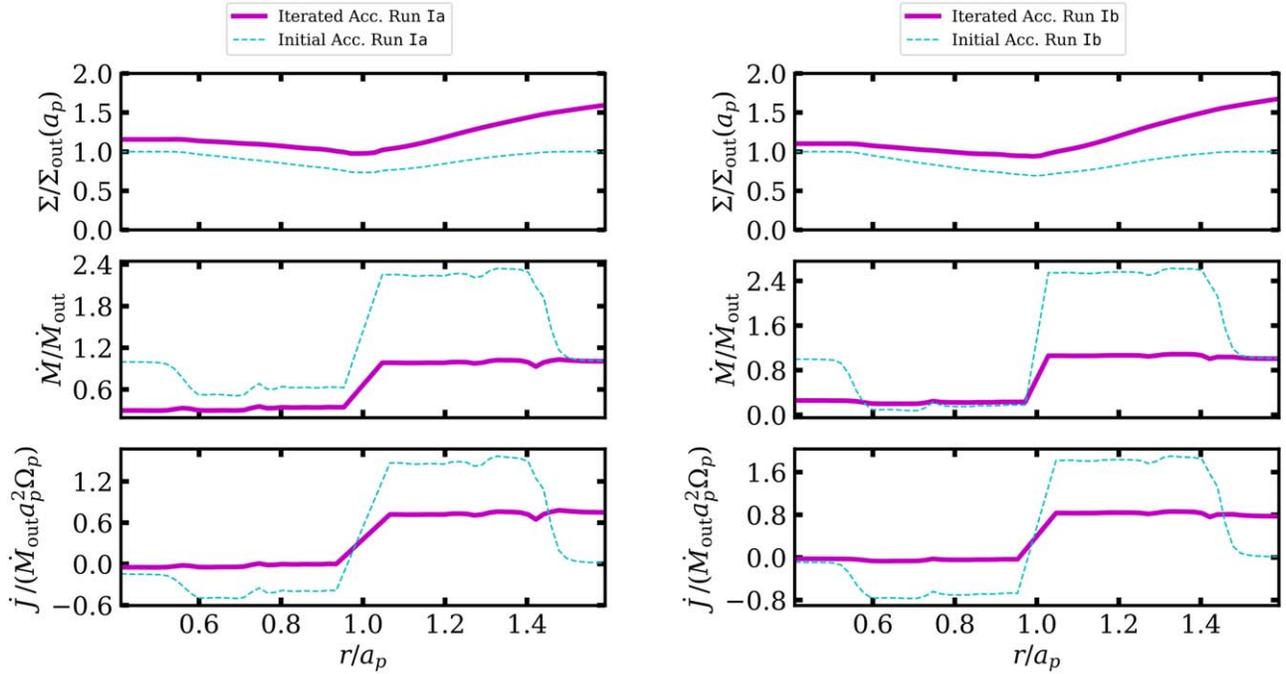

**Figure 6.** The top two panels are the same as Figure 4, but for Runs Ia (left) and Ib (right) (with sink radii different from Run I, $r_s = 0.05 r_H$ and $r_s = 0.2 r_H$, respectively). The $\dot{J}$ profiles before and after iteration are plotted in the bottom panels.

An accreting planet in Run I accretes at a rate $\dot{m}_p/\dot{M}_{\rm out} = 0.708$; for this run, $\Delta \dot{M}/\dot{M}_{\rm out} = 0.715$, i.e., they agree to within $\sim 1\%$. The AM exchange rate $\dot{J}_p/(\dot{M}_{\rm out} a_p^2 \Omega_p) = 0.748$ and the corresponding $\Delta \dot{J}/(\dot{M}_{\rm out} a_p^2 \Omega_p) = 0.799$; they agree to within $\sim 7\%$. For an accreting planet, $\dot{J}_{p,a}$ dominates $\dot{J}_p$, and the pressure and gravitational components are negligible. The accretion eigenvalue $l_p = \dot{J}_p/(\dot{m}_p a_p^2 \Omega_p) = 1.06 > 1$. Using Equation (15), we find orbital expansion at a rate $\dot{a}_p/a_p = 0.0788 \dot{M}_{\rm out}/m_p$. Moreover, the absolute value of the migration rate is about 4 times larger than for a nonaccreting planet (Run In). If this effect is general and robust, it would mean that accretion could dominate migration, throwing the canonical gravity-only calculations into question.

### 4.1. Sink Radius

We perform two additional experiments based on Run I to test the effects of the sink radius, one with half the fiducial $r_s$ (Run Ia) and the other with double (Run Ib). We plot the disk profiles in Figure 6. The accretion rate, $\dot{m}_p$, and AM exchange rate, $\dot{J}_p$, decrease in Run Ia and increase in Run Ib. However, the orbital expansion rate increases to $\dot{a}_p/[(\dot{M}_{\rm out}/m_p) a_p] = 0.106$ (from 0.0788) in Ia and decreases to $\dot{a}_p/[(\dot{M}_{\rm out}/m_p) a_p] = 0.0276$ in Ib. Most importantly, changing $r_s$ does not change the sign of migration for the planet masses and sink radii explored here. Run Ia is the most realistic case due to its small sink radius. In this run, the smaller planetary accretion rate offsets the smaller value of $\dot{J}_p$, leading to the larger orbital expansion rate.

### 4.2. Planet Mass

In Run II we decrease the planet mass ratio to $q = 10^{-4}$. Run II has the smallest orbital expansion rate of all three accreting simulations, $\dot{a}_p/a_p = 0.0121 \dot{M}_{\rm out}/m_p$. However, in this case the absolute value of the migration rate is $\sim 16\times$ larger than in Run IIn.

The accretion rate and torque shrink to $\dot{m}_p/\dot{M}_{\rm out} = 0.488$ and $\dot{J}_p/(\dot{M}_{\rm out} a_p^2 \Omega_p) = 0.494$, respectively. We plot the disk profiles in Figure 7. Compared to Runs I, Ia, and Ib, the disk is largely unperturbed except for the steps in $\dot{M}(r)$ and $\dot{J}(r)$ at $r = a_p$.

### 5. Summary and Discussion

We have performed 2D global hydrodynamic simulations to investigate the coupled effects of accretion and migration for a protoplanet embedded in a PPD (or a stellar-mass BH embedded in an AGN disk). Treating the accreting planet (in a fixed circular orbit) as an absorbing sphere, we compute the mass accretion ($\dot{m}_p$) and torque on the planet ($\dot{J}_p$) directly, including all forms of AM exchange between the planet in the disk, through accretion, gravity, pressure, and viscosity. We develop an iterative scheme so that our simulated disk-planet system reaches a steady state: The disk mass and AM fluxes are constant (independent of $r$) except for jumps ($\Delta \dot{M}$ and $\Delta \dot{J}$) at the planet's orbital radius, and these jumps match the planetary mass-accretion rate ($\dot{m}_p = \Delta \dot{M}$) and torque on the planet ($\dot{J}_p = \Delta \dot{J}$).

Figure 8 summarizes our results for the planets' accretion rates, AM exchange rates, and orbital expansion/contraction rates (see also Table 1). All accreting runs have a numerically determined accretion eigenvalue $l_p > 1$, computed either as

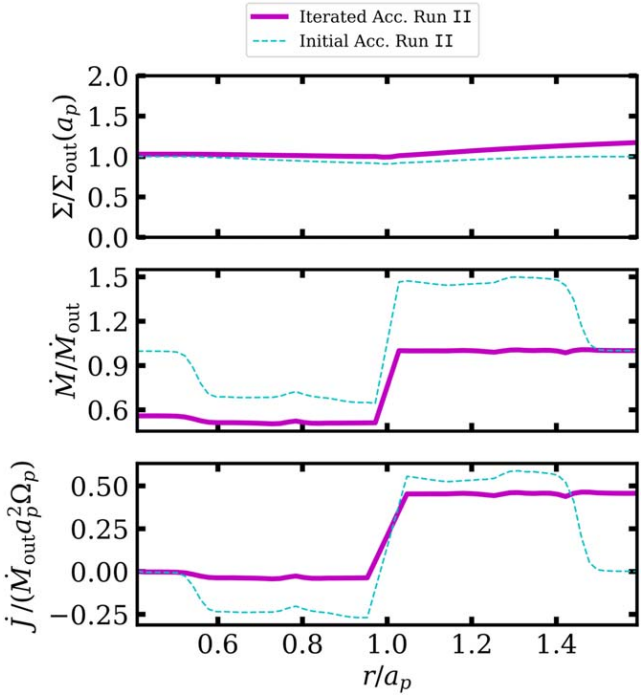

**Figure 7.** Same as Figure 6, but for Run II.

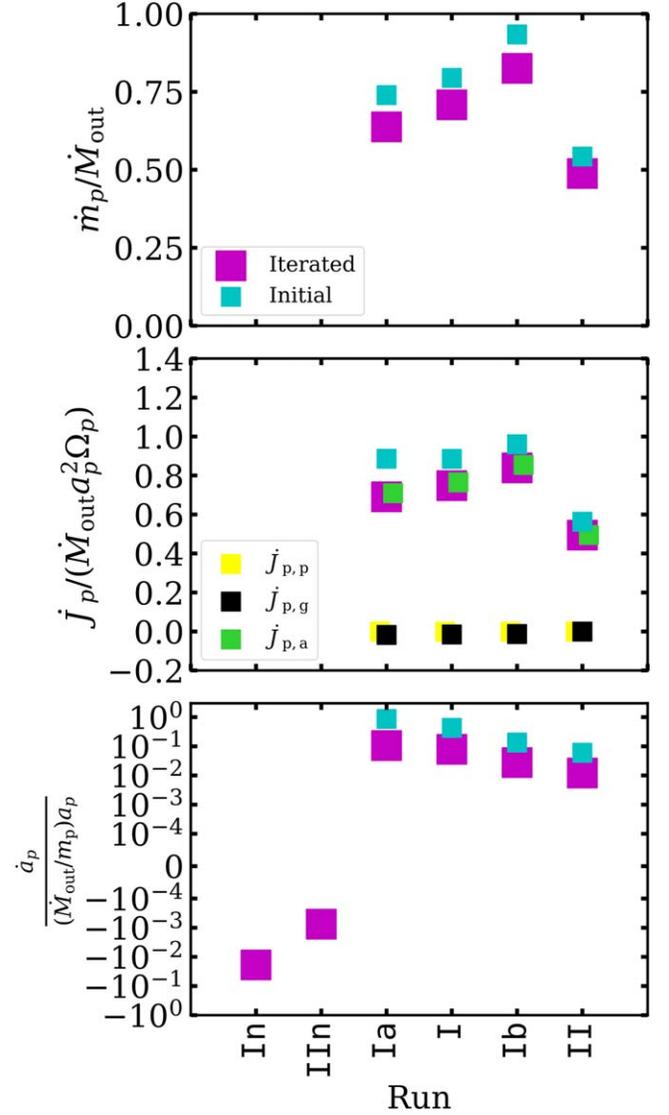

**Figure 8.** The key results of this study. The top two plots are only for the accreting planets. The bottom plot includes results for nonaccreting planets as well. Magenta squares indicate measurements from the iterated simulations (see Section 2.4) and cyan squares indicate measurements from before iteration. Top: the planetary accretion rates $\dot{m}_p$. Middle: the torques on the planets $\dot{J}_p$. The magenta and cyan squares are total AM exchange, $\dot{J}_p = \dot{J}_{p,p} + \dot{J}_{p,g} + \dot{J}_{p,a}$, and the yellow, black, and green squares the components. Bottom: the orbital expansion/contraction rate, $\dot{a}_p$.

$l_p \equiv \dot{J}_p/(\dot{m}_p a_p^2 \Omega_p)$ or $l_p = \Delta \dot{J}/(\Delta \dot{M} a_p^2 \Omega_p)$. The two different ways of calculating $l_p$ provide independent checks for our results. This finding means that an accreting planet migrates outward (Equation (15)), which is our most important result. Furthermore, in our fiducial model of accreting planets (Run I), the magnitude of the (outward) orbital migration rate is several times larger than that of a nonaccreting planet (migrating inward).

Our results come with several important caveats. First, we are utilizing a 2D model. While our results are self-consistent within the 2D setup, our disk model has a scale height $H$ comparable to the planet/BH Hill radius $r_H$. This means that our 2D simulations can describe 3D flows only approximately. Three-dimensional models which take vertical gas flow into account are the only way to test the robustness of our results. Three-dimensional effects are particularly significant for smaller planet/BH masses, for which $r_H < H$. This is an important direction for future research.

Second, we have adopted an idealized "absorbing sphere" accretion prescription. There is an inherent uncertainty in the appropriate choice for the sink radius, $r_s$. Smaller sink radii may give more realistic results (given that they are still much larger than the accretor's size). The smallest we use in our study is 5% of $r_H$ (Run Ia). Part of this caveat is that there is a trend of increasingly positive $\dot{a}_p$ as we shrink $r_s$. Extending the investigation to smaller $r_s$ until $\dot{a}_p$ levels off could quantify convergence. The transition at the sink radius to effectively a vacuum is inherently unrealistic. Other authors have implemented the sink term in different ways, which give a smoother behavior near the accretor (e.g., W. Kley et al. 2001; G. D'Angelo et al. 2002, 2003; Y.-P. Li et al. 2023). Additionally, some authors have argued for a "torque-free" sink, which could also improve convergence (e.g., A. M. Dempsey et al. 2020; A. J. Dittmann & G. Ryan 2021; A. M. Dempsey et al. 2022). A more realistic model would take the thermal structure of the planet (or mini-disk around the BH) into account and evolve the disk-accretor system in a thermodynamically self-consistent manner. Just like the circumbinary accretion problem (see D. Lai & D. J. Muñoz 2023), where the binary evolution (inspiral or outspiral) depends on the disk thermodynamics, much work remains to be done for accreting planets/BHs in this direction.

Finally, our planet/BH is on a fixed, circular orbit. A more realistic simulation would have the accretor's orbit self-consistently evolve. Because the orbit is circular, we do not know how finite eccentricity might affect the planet's/BH's accretion, and we cannot use our results to calculate the eccentricity evolution of the planet/BH. We plan to address this issue in a future paper.

After the completion of this work, we noticed a similar work by Y.-P. Li et al. (2024), who find similar conclusions in 2D and 3D. They explore the viscosity parameter and find a similar result (outward migration) in the parameter space we explored (e.g., their model A1). In most of their simulations, they find a larger gravitational torque compared to the accretion torque, contrary to our results. This discrepancy is probably due to the different accretion prescriptions or from the fact that their simulations have not achieved full global steady state as in our simulations (e.g., see Figures 6 and 7).

## Acknowledgments

This work is funded in part by Cornell University. R.L. acknowledges support from the Heising-Simons Foundation 51 Pegasi b Fellowship. We would like to thank Ya-Ping Li, Martin Pessah, Wenrui Xu, and Zhaohuan Zhu for helpful discussions.

## ORCID iDs

JT Laune ● https://orcid.org/0000-0002-0896-7393
Rixin Li (李日新) ● https://orcid.org/0000-0001-9222-4367
Dong Lai ● https://orcid.org/0000-0002-1934-6250

## References

Artymowicz, P., Lin, D. N. C., & Wampler, E. J. 1993, ApJ, 409, 592
Ayliffe, B. A., & Bate, M. R. 2012, MNRAS, 427, 2597
Baruteau, C., & Masset, F. 2008, in IAU Symp. 249, Exoplanets: Detection, Formation and Dynamics, ed. Y.-S. Sun, S. Ferraz-Mello, & J.-L. Zhou (Cambridge: Cambridge Univ. Press), 397
Bellovary, J. M., Low, o. M., McKernan, B., & Ford, K. E. S. 2016, ApJ, 819, L17
Benítez-Llambay, P., Masset, F., Koenigsberger, G., & Szulágyi, J. 2015, Natur, 520, 63
Chametla, R. O., & Masset, F. S. 2021, MNRAS, 501, 24
Chrenko, O., Brož, M., & Lambrechts, M. 2017, A&A, 606, A114
D'Angelo, G., Henning, T., & Kley, W. 2002, A&A, 385, 647
D'Angelo, G., Kley, W., & Henning, T. 2003, ApJ, 586, 540
De Val-Borro, M., Edgar, R. G., Artymowicz, P., et al. 2006, MNRAS, 370, 529
Dempsey, A. M., Li, H., Mishra, B., & Li, S. 2022, CoRR, Astrophysics Source Code Library, ascl:2203.06534
Dempsey, A. M., Munoz, D., & Lithwick, Y. 2020, ApJL, 892, L29
Dittmann, A. J., & Ryan, G. 2021, ApJ, 921, 71
Dittmann, A. J., & Ryan, G. 2022, MNRAS, 513, 6158
Dittmann, A. J., & Ryan, G. 2024, ApJ, 967, 12
D'Orazio, D. J., & Duffell, P. C. 2021, ApJL, 914, L21
Duffell, P. C., D'Orazio, D., Derdzinski, A., et al. 2020, ApJ, 901, 25
Duffell, P. C., Dittmann, A. J., D'Orazio, D. J., et al. 2024, ApJ, 970, 156
Dürmann, C., & Kley, W. 2017, A&A, 598, A80
Eklund, H., & Masset, F. S. 2017, MNRAS, 469, 206
Goldreich, P., & Tremaine, S. 1979, ApJ, 233, 857
Goldreich, P., & Tremaine, S. 1980, ApJ, 241, 425
Hankla, A. M., Jiang, Y.-F., & Armitage, P. J. 2020, ApJ, 902, 50
Heath, R. M., & Nixon, C. J. 2020, A&A, 641, A64
Hori, Y., & Ikoma, M. 2011, MNRAS, 416, 1419
Ikoma, M., Nakazawa, K., & Emori, H. 2000, ApJ, 537, 1013
Jimenez, M. A., & Masset, F. S. 2017, MNRAS, 471, 4917
Kley, W., & Crida, A. 2008, A&A, 487, L9
Kley, W., D'Angelo, G., & Henning, T. 2001, ApJ, 547, 457
Kley, W., & Nelson, R. P. 2012, ARA&A, 50, 211
Lai, D., & Muñoz, D. J. 2023, ARA&A, 61, 517
Lega, E., Crida, A., Bitsch, B., & Morbidelli, A. 2014, MNRAS, 440, 683
Levin, Y. 2007, MNRAS, 374, 515
Li, J., Lai, D., & Rodet, L. 2022, ApJ, 934, 154
Li, R., & Lai, D. 2022, MNRAS, 517, 1602
Li, R., & Lai, D. 2023, MNRAS, 522, 1881
Li, R., & Lai, D. 2024, MNRAS, 529, 348
Li, Y.-P., Chen, Y.-X., & Lin, D. N. C. 2023, MNRAS, 526, 5346
Li, Y.-P., Chen, Y.-X., & Lin, D. N. C. 2024, ApJ, 971, 130
Lin, D. N. C., & Papaloizou, J. 1979, MNRAS, 186, 799
Masset, F. S. 2001, ApJ, 558, 453
Masset, F. S. 2017, MNRAS, 472, 4204
Masset, F. S., & Casoli, J. 2009, ApJ, 703, 857
Masset, F. S., & Casoli, J. 2010, ApJ, 723, 1393
Masset, F. S., & Ogilvie, G. I. 2004, ApJ, 615, 1000
McKernan, B., Ford, K. E. S., Lyra, W., & Perets, H. B. 2012, MNRAS, 425, 460
Miranda, R., Munoz, D. J., & Lai, D. 2017, MNRAS, 466, 1170
Moody, M. S. L., Shi, J.-M., & Stone, J. M. 2019, ApJ, 875, 66
Muñoz, D. J., Lai, D., Kratter, K., & Miranda, R. 2020, ApJ, 889, 114
Munoz, D. J., Miranda, R., & Lai, D. 2019, ApJ, 871, 84
Ormel, C. W. 2013, MNRAS, 428, 3526
Paardekooper, S.-J., Baruteau, C., Crida, A., & Kley, W. 2010, MNRAS, 401, 1950
Paardekooper, S.-J., Baruteau, C., & Kley, W. 2011, MNRAS, 410, 293
Paardekooper, S.-J., Dong, R., Duffell, P., et al. 2023, in ASP Conf. Ser. 534, Protostars and Planets VII, ed. S. Inutsuka (San Francisco, CA: ASP), 685
Paardekooper, S.-J., & Mellema, G. 2006, A&A, 459, L17
Paardekooper, S.-J., & Papaloizou, J. C. B. 2008, A&A, 485, 877
Paardekooper, S. J., & Papaloizou, J. C. B. 2009a, MNRAS, 394, 2283
Paardekooper, S. J., & Papaloizou, J. C. B. 2009b, MNRAS, 394, 2297
Piso, A.-M. A., & Youdin, A. N. 2014, ApJ, 786, 21
Pollack, J. B., Hubickyj, O., Bodenheimer, P., et al. 1996, Icar, 124, 62
Rafikov, R. R. 2006, ApJ, 648, 666
Secunda, A., Bellovary, J., Low, o. M., et al. 2019, ApJ, 878, 85
Stone, J. M., Tomida, K., White, C. J., & Felker, K. G. 2020, ApJS, 249, 4
Tagawa, H., Haiman, Z., & Kocsis, B. 2020, ApJ, 898, 25
Tanaka, H., Takeuchi, T., & Ward, W. R. 2002, ApJ, 565, 1257
Wang, H.-Y., Bai, X.-N., Lai, D., & Lin, D. N. C. 2023, MNRAS, 526, 3570
Ward, W. R. 1986, Icar, 67, 164
Ward, W. R. 1997, Icar, 126, 261
Zrake, J., Tiede, C., MacFadyen, A., & Haiman, Z. 2021, ApJL, 909, L13